\documentclass[12pt]{article}
\usepackage{epsfig}
\usepackage{url}
\usepackage{hyperref}
\def\beq{\begin{equation}}
\def\eeq#1{\label{#1}\end{equation}}
\def\eeqn{\end{equation}}

\def\beqa{\begin{eqnarray}}
\def\eeqa#1{\label{#1}\end{eqnarray}}
\def\eeqan{\end{eqnarray}}

\let\bar=\overbar

\def\Dslash{\not{\hbox{\kern-4pt $D$}}}
\def\dslash{\not{\hbox{\kern-2pt $\del$}}}

\def\msb{{\bar{\ssstyle M \kern -1pt S}}}

 \def\preprint#1{\begin{flushright} {\tt #1 } \end{flushright}}
\def\Title#1{\begin{center} {\Large {\bf #1} } \end{center}}

\begin{document}

\preprint{IFT-UAM/CSIC-11-71\\UAB-FT-697}
\Title{Anomalies and Transport Coefficients: The Chiral Gravito-Magnetic Effect}

\bigskip\bigskip


\begin{raggedright}  

{\it{ \bf Karl Landsteiner$^{a}$, Eugenio Meg\'\i as$^{a,b}$ and
    Francisco Pena-Benitez$^{a,c}$\index{Landsteiner, K.}}\\ 
  $^a$Instituto de F\'\i sica Te\'orica UAM-CSIC,
  Universidad Aut\'onoma de Madrid,\\ Cantoblanco E-28049 Madrid,
  Spain\\ 
  $^{b}$Grup de F\'{\i}sica Te\`orica and IFAE, Departament de
  F\'{\i}sica, Universitat Aut\`onoma de Barcelona, Bellaterra E-08193
  Barcelona, Spain\\ 
  $^{c}$Departamento de F\'{\i}sica Te\'orica,
  Universidad Aut\'onoma de Madrid,\\ Cantoblanco E-28049 Madrid,
  Spain}
\bigskip\bigskip
\end{raggedright}

\section{Introduction}

At finite temperature and density
anomalies give rise to new non-dissipative transport phenomena in the
hydrodynamics of charged relativistic fluids 
\cite{Fukushima:2008xe,Erdmenger:2008rm,Banerjee:2008th,Son:2009tf}.
As is beautifully explained in the RHIC-made video \cite{Brookhaven:2010:Online}
magnetic fields and vortices in the fluid induce currents via the
so-called chiral magnetic and chiral vortical conductivities. Although
there have been many early precursors that found manifestations of
this phenomena in the physics of neutrinos
\cite{Vilenkin:1995um,Vilenkin:1980ft,Vilenkin:1980zv,Vilenkin:1978hb},
the early universe \cite{Giovannini:1997eg} and condensed matter systems
\cite{Alekseev:1998ds}, the recent surge of interest is clearly
related to the physics of the quark gluon plasma.  It has been
suggested that the observed charge separation in heavy ion collisions
is related to a particular manifestation of these anomalous transport
phenomena: the chiral magnetic effect \cite{Fukushima:2008xe,Kharzeev:2007jp}. The
latter describes how a (electro-magnetic) B-field induces
via the axial anomaly an electric current parallel to the magnetic
field.  The first application of holography to the anomalous
hydrodynamics is \cite{Newman:2005hd} where the anomalous transport
effects due to R-charge magnetic fields have been examined.  Later
studies showed that there is also a related vortical effect
\cite{Erdmenger:2008rm,Banerjee:2008th}, i.e. a vortex in the fluid
induces a current parallel to the axial vorticity vector $\Omega^\mu =
\epsilon^{\mu\nu\rho\lambda} u_\nu \partial_\rho u_\lambda$, and
related effects of the presence of angular momentum had been discussed
before in a purely field theoretical setup in \cite{Vilenkin:1980zv}
and \cite{Kharzeev:2007tn}.  
Studies of the chiral magnetic effect
using holography have appeared in
\cite{Yee:2009vw,Torabian:2009qk,Rebhan:2009vc,Brits:2010pw,Gynther:2010ed,Gorsky:2010xu,Amado:2011zx,Kalaydzhyan:2011vx,Hoyos:2011us,Eling:2010hu}
and using lattice field theory in
\cite{Buividovich:2009wi,Abramczyk:2009gb,Yamamoto:2011gk}. 
The experimental status of the observed charge separation in heavy ion collision is
discussed in \cite{abelev:2009uh,abelev:2009txa}. A related
effect is the so called chiral separation effect that induces an axial
current in a magnetic field \cite{Newman:2005as}. It has been argued
to lead to enhanced production of high spin hadrons in \cite{KerenZur:2010zw}.
In \cite{Son:2009tf} the authors showed that purely hydrodynamic considerations based on
demanding a positive definite divergence of the entropy current fix the
chiral magnetic and chiral vortical conductivities almost uniquely. In \cite{Neiman:2010zi}
it was however pointed out that there are ambiguities stemming from integration
constants that allow additional temperature dependence. It is precisely this
temperature dependence that we can fix by calculating the transport coefficients
via Green-Kubo formulas and discover a somewhat surprising relation to the gravitational anomaly.

Anomalies belong to the most interesting and most subtle properties of
relativistic quantum field theories.  They are responsible for the
breakdown of a classical symmetry due to quantum effects. The
Adler-Bardeen non-renormalization theorem guarantees that this
breakdown is saturated at the one-loop level.  Therefore the presence
of anomalies can be determined through simple algebraic criteria on
the representations under which the chiral fermions of a particular
theory transform.  In vacuum the anomaly appears as the
non-conservation of a classically conserved current in a triangle
diagram with two additional currents. In four dimension two types of
anomalies can be distinguished according to whether only spin one
currents appear in the triangle \cite{Adler:1969gk,Bell:1969ts} or if also the
energy-momentum tensor participates \cite{Delbourgo:1972xb,Eguchi:1976db}. We will
call the first type of anomalies simply chiral anomalies and the
second type gravitational anomalies. To be precise, in four dimension
we should actually talk of mixed gauge-gravitational anomalies since
triangle diagrams with only energy-momentum insertions are perfectly
conserved (see
e.g. \cite{AlvarezGaume:1983ig}). In four dimensional Minkowski space massless fermions 
can always be written in a 
basis of only left-handed fermions. If they transform in a representation $T_A$ of a symmetry 
the presence of chiral anomalies
is detected by the non-vanishing of $d_{ABC} = \frac 1
2\mathrm{Tr}(T_A\{T_B,T_C\})$ whereas the presence of a gravitational
anomaly is detected by the non-vanishing of $b_A= \mathrm{Tr}(T_A)$.

\section{Kubo Formulas}

Transport coefficients can be computed in linear response theory via so-called Green-Kubo formulas. 
The response of a system in equilibrium to an external perturbation is encoded in the retarded 
Green's functions. If we apply an electric field to our system we will generate an electric current.
If the electric field is (sufficiently) weak the magnitude of the current will be proportional to
the applied electric field and the constant of proportionality is simply the electric conductivity
$\sigma^{\cal E}$.
\begin{equation}
 \vec{J} = \sigma^{\cal E} \vec{\cal E}\,.
\end{equation}
We can derive a Green-Kubo formula by noting that the electric field is $\vec{\cal E} = -i \omega \vec{\cal A}$ in terms
of a vector potential $\vec{A}$. We now interpret above equation as an expression for the vacuum expectation
value of the current in the background of an external gauge field $\vec{\cal A}$. Since the vector potential acts
as a source for the current we can functionally differentiate with respect to it and get an expression for the
(retarded) two-point function of currents. In particular we can obtain in this way the Green-Kubo formula
for the DC-conductivity
\begin{equation}
 \sigma^{\cal E} = \lim_{\omega\rightarrow 0} \frac{i}{\omega} \langle J_i J_i \rangle|_{\vec{k}=0}\,,
\end{equation}
(no summation over the spatial index $i$ here). 

Let us now jump to the chiral magnetic effect. It describes the generation of a current due to a magnetic
field, so
\begin{equation}
 \vec{J} = \sigma^{\cal B} \vec{\cal B}\,.
\end{equation}
Proceeding as before it is straightforward to see that the Green-Kubo formula for the chiral magnetic conductivity
$\sigma^{\cal B}$ is
\begin{equation}
 \sigma^{\cal B} =  \lim_{k_n\rightarrow 0} \sum_{i,j}\epsilon_{ijn} \frac{i}{2 k_n} \langle J_i J_j \rangle|_{\omega=0}\,.
\end{equation}
So far we have only discussed the response to electro-magnetic $\cal E$- and $\cal B$-fields. From now on we want to
be a bit more general and imagine to have an arbitrary symmetry group with generators $T_A$ and couple them
to non-dynamical, i.e. external gauge fields ${\cal A}_A$ such that their variation inserts the currents $J_A$ into
correlation functions. The obvious generalization of the chiral magnetic conductivity is then
\begin{equation}
 \sigma^{\cal B}_{AB} = \lim_{k_n\rightarrow 0} \sum_{i,j}\epsilon_{ijn} \frac{i}{2 k_n} \langle J^i_A J^j_B \rangle|_{\omega=0}\,.
\end{equation}
That is $\sigma^{\cal B}_{AB}$ describes the generation of the current $\vec J_A$ if an external field $\vec{\cal B}_B = \vec{\nabla}\times \vec{\cal A}_B$
is switched on\footnote{In principle we should include also commutator terms in ${\cal B}_B$. They are however not important if we
want to study only two point functions of currents with vanishing external fields.}. 

If the fluid under consideration is charged with respect to some of the (classically) conserved charges there will necessarily be
an energy transport related to the charge transport induced by the current $\vec{J}_A$. We know that a variation in the
charge distribution will cost us an energy of the form $\delta \epsilon = \mu_A \delta Q_A$. If we imagine a test charge
$\delta Q_A$ moving through the charged plasma it will therefore generate a current $\delta \vec{J}_A$ and induce also an energy
current of the form $\delta T^{0i} = \mu_A \delta J^i_A$. For a finite current we should integrate this over $\mu_A$
and obtain $T^{0i} = \int \mu'_A  \frac{d J^i}{d\mu'_A} d\mu'_A$. It follows therefore that the energy flux due
to an external ${\cal B}_A$ field is measured by the transport coefficient 
\begin{equation}
 \sigma^{\cal B}_B =  \lim_{k_n\rightarrow 0} \sum_{i,j}\epsilon_{ijn} \frac{i}{2 k_n} \langle T^{0i} J_B^j \rangle|_{\omega=0}
= \int \mu_A d\sigma^{\cal B}_{A B} + \mathrm{const.} \,.
\end{equation}
We explicitly introduced a undetermined integration constant here. Evaluating the Kubo formula we will see that the
integrations constant is $\propto T^2$ and non-zero only if gravitational anomalies are present.

The correlators are retarded ones but evaluated at zero frequency. 
For this reason the order of the operators can be reversed
and we we also can define
\begin{equation}
 \sigma^{\cal V}_A = \lim_{k_n\rightarrow 0} \sum_{i,j}\epsilon_{ijn} \frac{i}{2 k_n} \langle J_A^i T^{0j} \rangle|_{\omega=0}
\end{equation}
Although it is clear that $\sigma^{\cal B}_A = \sigma^{\cal V}_A$ they describe different transport phenomena. Whereas
$\sigma^{\cal B}_A$ describes the generation of an energy flux due to an external ${\cal B}_A$ field, $\sigma^{\cal V}_A$
describes the generation of the current $\vec{J}_A$ due to an external field that sources $T^{0i}$. It is not difficult
to convinces oneself that the field in question is the so-called gravito-magnetic field defined by a variation of the
(flat) metric of the form
\begin{equation}
 ds^2 = -dt^2 + 2 \vec{\cal A}_g dt d\vec{x} + d\vec{x}^2\,.
\end{equation}
If we linearize gravity with this metric $\vec{\cal A}_g$ acts indeed like a normal abelian vector potential. We might therefore
call the new transport coefficient $\sigma^{\cal V}_A$ a ``chiral gravito-magnetic'' conductivity giving rise to a ``chiral gravito-magnetic effect''. In a fluid a second, and somewhat
more down-to earth interpretation is available. The flow of a relativistic fluid is characterized by the fluid four-velocity $u^\mu$. In the restframe
of the fluid but in the background of a gravito-magnetic potential we have $u_\mu = (-1,\vec{v})=(-1,\vec{\cal A})$. Therefore the gravito-magnetic field can also be calculated as the curl of the velocity field  ${\cal \vec{B}}_g  = \vec{\nabla}\times \vec{v}$.
It is therefore natural to interpret $\sigma^{\cal V}_A$ as the response in the current due to a vortex in the fluid, i.e. as
a ``chiral vortical conductivity''. 

Since now we have convinced ourselves that vortices (or gravito-magnetic fields) in the fluid might generate currents it comes
as no surprise that they also will generate an energy flux and that the corresponding conductivity
will be given by
\begin{equation}
 \sigma^{V} = \lim_{k_n\rightarrow 0} \sum_{i,j}\epsilon_{ijn} \frac{i}{2 k_n} \langle T^{0i} T^{0j} \rangle|_{\omega=0}
\end{equation}

In a hydrodynamic framework we can summarize our findings in the constitutive relations
\begin{eqnarray}
 T^{\mu\nu}  &=& (\epsilon+P) u^\mu u^\nu + P \eta^{\mu\nu} + Q^\mu u^\nu + Q^\nu u^\mu \,,\\
 J^\mu_A &=& n_A u^\mu + N^\mu_A\,.
\end{eqnarray}
with the first order in derivatives terms
\begin{eqnarray}
 N^\mu_A &=& \sigma^{\cal B}_{AB} {\cal B}^\mu_B + \sigma^{\cal V}_{A} \Omega^\mu \,,\\
Q^\mu &=& \sigma^{\cal B}_{A} {\cal B}^\mu_A + \sigma^{\cal V}_{A} \Omega^\mu\,.
\end{eqnarray}
For simplicity of the expressions we have dropped here the usual dissipative terms related to shear and bulk viscosities or
electric conductivity.
The equilibrium quantities $\epsilon, P, n_A$ are energy density, pressure and charge densities and we
defined the (covariant) magnetic fields ${\cal B}_A^\mu = \epsilon^{\mu\nu\rho\lambda} u_\nu \partial_\rho A_{\lambda,A}$ and vorticity vector $\Omega^\mu = \epsilon^{\mu\nu\rho\lambda} u_\nu \partial_\rho u_\lambda$. 
The gravito-magnetic Lorentz force on a test particle of mass $m$ is \footnote{See \cite{Mashhoon:2003ax} but note
that our definition of the gravito-magnetic potential differs by a factor of $2$.}
\begin{equation}
 \vec{F} = m.\vec{v}\times \vec{\cal B}_g\,.
\end{equation}
It follows that the work done by the gravito-magnetic field on a test particle is $W=\int \vec{F} .\vec{v} dt =0$ just as in the case of a usual magnetic field.
Since neither a magnetic nor a gravito-magnetic field (or a vortex) do work on the system these constitutive
relations describe dissipationless transport. This property and the related T-invariance of the transport have recently
been emphasized in \cite{Loganayagam:2011mu, Kharzeev:2011ds} where these constitutive relations have been generalized to higher dimensions as well.

We also note that the fluid velocity in the hydrodynamic derivative expansion suffers from ambiguities. Indeed we can always
redefine $u^\mu\rightarrow u^\mu+\delta u^\mu$ and declare $\delta u^\mu$ to be of the same order as $Q^\mu$ and $N^\mu$.
Choosing $\delta u^\mu = -Q^\mu/(\epsilon+P)$ effectively removes $Q^\mu$ from the constitutive relation of the energy-momentum
tensor and defines the so called Landau frame. The current is then given by
\begin{equation}
 N^\mu = \xi^{\cal B}_{A B} {\cal B}_B + \xi^{\cal V}_{A} \Omega^\mu
\end{equation}
with the Landau frame transport coefficients
\begin{eqnarray}
 \xi^{\cal B}_{AB} &=& \sigma_{AB}^{\cal B} - \frac{n_A}{\epsilon + P} \sigma^{\cal B}_{B}\,,\\
 \xi^{\cal V}_{A} &=& \sigma_{A}^{\cal V} - \frac{n_A}{\epsilon + P} \sigma^{\cal V}\,,
\end{eqnarray}

\subsection{Weak coupling}
Our aim is now to evaluate these Green-Kubo's formulas in a theory of free 
right-handed fermions $\Psi^f$ transforming under a
global symmetry group $G$ generated by matrices $(T_A)^f\,_g$.  We
denote the generators in the Cartan subalgebra by $H_A$. Chemical
potentials $\mu_A$ can be switched on only in the Cartan
subalgebra. Furthermore the presence of the chemical potentials breaks
the group $G$ to a subgroup $\hat G$. Only the currents that lie in
the unbroken subgroup are conserved (up to anomalies) and participate
in the hydrodynamics. The chemical potential for the fermion $\Psi^f$
is given by $\mu^f= \sum_A q_A^f  \mu_A$, where we write the Cartan generator
$H_A = q_A^f\delta^f\,_g$ in terms of its eigenvalues, the charges
$q_A^f$. The unbroken symmetry group $\hat G$ is generated by those
matrices $T_A^f\,_g$ fulfilling
\begin{equation}\label{eq:unbroken}
 T_A^f\,_g \mu^g = \mu^f T_A^f\,_g\,.
\end{equation}
There is no summation over indices in the last expression. From now on we will assume that all currents $\vec{J}_A$ lie in directions indicated in  (\ref{eq:unbroken}). We define the chemical potential through boundary conditions on the fermion fields around the thermal circle \cite{Landsman:1986uw}, $ \Psi^f(\tau) = - e^{\beta \mu^f} \Psi^f(\tau-\beta)$ with $\beta=1/T$. 
Therefore the eigenvalues of $\partial_\tau$ are
$i\tilde\omega_n+\mu^f$ for the fermion species $f$ with
$\tilde\omega_n=\pi T(2n+1)$ the fermionic Matsubara frequencies.  A
convenient way of expressing the currents is in terms of Dirac fermions
and writing
\begin{eqnarray}
J^i_A &=& \sum_{f,g=1}^N T_A^g\,_f \bar\Psi_g \gamma^i  {\cal P}_+ \Psi^f \,, \label{eq:JA}\\
T^{0i} &=&  \frac i 2 \sum_{f=1}^N\bar\Psi_f  ( \gamma^0  \partial^i + \gamma^i \partial^0  ) {\cal P}_+\Psi^f\,, \label{eq:JE}
\end{eqnarray}
where we used the chiral projector ${\cal P}_\pm = \frac 1 2 (1\pm\gamma_5)$.
The fermion propagator is
\begin{eqnarray}
S(q)^f\,_g &=&  \frac{\delta^f\,_g}{2} \sum_{t=\pm} \Delta_t(i\tilde\omega^f,\vec{q}) {\cal P}_+ \gamma_\mu \hat q^\mu_t \,,\\
\Delta_t( i\tilde\omega^f, q) &=& \frac{1}{i\tilde\omega^f - t E_q}\,,
\end{eqnarray}
with  $i\tilde\omega^f = i\tilde\omega_n + \mu^f$, $\hat q_t^\mu = (1, t \hat q)$, $\hat{q} = \frac{\vec{q}}{E_q}$ and $E_q=|\vec q |$. 
We can easily include left-handed fermions as well. 

The relevant Green's function for the chiral magnetic can be evaluated with standard finite temperature techniques (see~\cite{Landsteiner:2011cp} for details).
The results for the different conductivities are neatly summarized as
\begin{eqnarray}\label{eq:cmc}
\sigma^{\cal B}_{AB} &=& \frac{1}{4 \pi^2} d_{ABC}\mu^C\,,\\ \label{eq:cvc}
\sigma^{\cal V}_{A} &=& \frac{1}{8\pi^2} d_{ABC} \mu^B\mu^C  + \frac{T^2}{24} b_A\,,\\ \label{eq:cvce}
\sigma^{\cal V} &=& \frac{1}{12\pi^2} d_{ABC}\mu^A\mu^B\mu^C + \frac{T^2}{12} b_A\mu^A\,.
\end{eqnarray}
The result shows that these conductivities are non-zero if and only if the theory features anomalies. 
Let us come back now to the above mentioned integration constants. We see that in they are fixed to 
a particular form proportional to $T^2$ with a coefficient that coincides with the gravitational anomaly coefficient.
Additional terms of the form $T^3$ are allowed on dimensional grounds in $\sigma^{\cal V} $ \cite{Neiman:2010zi}, CPT 
invariance forbids these terms however and indeed they don't show up in the Green-Kubo formulas.

In vacuum the anomaly appears on the level of three point functions. In the presence of external sources for the energy momentum tensor
and the currents this is conveniently expressed through \cite{Delbourgo:1972xb,Eguchi:1976db,AlvarezGaume:1983ig}
\begin{equation}
 \nabla_\mu J_A^\mu= \epsilon^{\mu\nu\rho\lambda}\left( \frac{d_{ABC}}{32\pi^2}  F^B_{\mu\nu} F^C_{\rho\lambda} + \frac{b_A}{768\pi^2}  R^\alpha\,_{\beta\mu\nu}
R^\beta\,_{\alpha\rho\lambda}\right) \,.\label{eq:anomaly}
\end{equation}
Note that the gravitational anomaly is actually fourth order in derivatives and therefore a naive counting argument would suggest
that it can not contribute to first order hydrodynamics. It is however more useful to count derivatives acting on the connections ${\cal A}^\mu_a$ and $\Gamma_{\mu\nu}^\rho$. In this way the Riemann tensor is first order in derivatives just like the gauge field strength.
Since this transport coefficients seem to be intimately related to anomalies it is tempting to speculate that they are not renormalized
as we switch on interactions. We will investigate this question in the next section.

\subsection{Strong coupling}

As is well known gauge theories at infinitely strong coupling can be investigated through the gauge-gravity correspondence. It is 
therefore interesting to see if the anomalous conductivities can be obtained also from a strong coupling calculations based on
the action 
\begin{eqnarray}\nonumber
S &=& \frac{1}{16\pi G} \int d^5x \sqrt{-g} \left[ R + 2 \Lambda -
  \frac 1 4 F_{MN} F^{MN} \right.\\ &&\left.+ \epsilon^{MNPQR} A_M
  \left( \frac\kappa 3 F_{NP} F_{QR} + \lambda R^A\,_{BNP} R^B\,_{AQR}
  \right) \right] + S_{GH} + S_{CSK} \,,\\ S_{GH} &=& \frac{1}{8\pi G}
\int_\partial d^4x \sqrt{-h} \, K \,,\\ S_{CSK} &=& - \frac{1}{2\pi G}
\int_\partial d^4x \sqrt{-h} \, \lambda n_M \epsilon^{MNPQR} A_N
K_{PL} D_Q K_R^L \,, 
\end{eqnarray} where $S_{GH}$ is the usual Gibbons-Hawking
boundary term and $D_A=h_A^B\nabla_B$ is the covariant derivative on
the four dimensional boundary.
The most important fact about this action is the presence of two Chern-Simons terms, one a pure gauge field CS term and the
second one a mixed gauge-gravitational CS-term. The action depends therefore explicitly on the gauge field $A_M$ and is
gauge invariant only up to a boundary term of the form
\begin{equation}
 \delta S = \frac{1}{16 \pi G} \int_\partial d^4 x \sqrt{-h} \epsilon^{mnkl}\left( \frac{\kappa}{3} \hat{F}_{mn} \hat{F}_{kl} + \lambda \hat R^{i}\,_{jmn} \hat R^{j}\,_{ikl}\right) \,.
\end{equation}
where $\hat R$ is the induced curvature on the boundary. This is indeed of the form of the anomaly and we can match the CS couplings
to the anomaly coefficients
\begin{eqnarray}
-\frac{\kappa}{48 \pi G} &=& \frac{b}{96 \pi^2}\,,\\
-\frac{\lambda}{16\pi G} &=& \frac{d}{768 \pi^2} \,.
\end{eqnarray}
In this simple model there is only one $U(1)$ symmetry and therefore the anomaly coefficients are simply $b=\sum q_i$ and $d=\sum q_I^3$.
where $q_i$ are the charges of the chiral fermions of the dual field theory. 
 
As background we chose the charged AdS black brane solution with metric and gauge field
\begin{eqnarray}
 ds^2 &=& r^2\left( -f(r) dt^2 + d\vec{x}^2\right) +\frac{r^2}{f(r)r^2}~~,~~f(r)=\left(1-\frac{M}{r^4}+\frac{Q^2}{r^6}\right)\,,\\
 A &=& \left(\beta -\frac{\mu r_H^2}{r^2}\right) dt\,.
\end{eqnarray}
The chemical potential is related to the charge  by $\mu = \sqrt{3} Q/r_H^2$ and is defined as the energy needed
to bring a unit charge from the boundary behind the horizon. The background value of the temporal component of
the boundary gauge field $\beta$ has to be distinguished from it along the lines explained in \cite{Gynther:2010ed}.
 
 Standard gauge-gravity techniques \cite{Son:2002sd, Kaminski:2009dh, Amado:2009ts} can now be employed in order to
 calculate the relevant Green's functions. It is straightforward
 although slightly tedious due to the complexities of the
 gauge-gravitational CS-term (see~\cite{Landsteiner:2011iq}). In any case it is reassuring to find
 precisely the form of (\ref{eq:cvc})-(\ref{eq:cvce}) specialized to the
 case of a single $U(1)$ theory. This can be
 taken as a strong hint pointing towards a non-renormalization of the
 anomalous conductivities.

\bigskip
{\bf Acknowledgements}: This work has been
supported by Plan Nacional de Altas Energ\'{\i}as FPA2009-07908, CPAN
(CSD2007-00042), Comunidad de Madrid HEP-HACOS S2009/ESP-1473. The research of E.M. is supported by the Juan de la Cierva 
Program of the Spanish MICINN.

\bibliographystyle{hunsrt}
\bibliography{Landsteiner}

\end{document}